\documentstyle{mn}
\newif\ifAMStwofonts
\AMStwofontstrue
\input psfig.tex
\ifoldfss
  \ifCUPmtlplainloaded \else
    \NewTextAlphabet{textbfit} {cmbxti10} {}
    \NewTextAlphabet{textbfss} {cmssbx10} {}
    \NewMathAlphabet{mathbfit} {cmbxti10} {} % for math mode
    \NewMathAlphabet{mathbfss} {cmssbx10} {} %  "   "    "
  \fi
  \ifAMStwofonts
    \ifCUPmtlplainloaded \else
      \NewSymbolFont{upmath} {eurm10}
      \NewSymbolFont{AMSa} {msam10}
      \NewMathSymbol{\upi}     {0}{upmath}{19}
      \NewMathSymbol{\umu}     {0}{upmath}{16}
      \NewMathSymbol{\upartial}{0}{upmath}{40}
      \NewMathSymbol{\leqslant}{3}{AMSa}{36}
      \NewMathSymbol{\geqslant}{3}{AMSa}{3E}

    \fi
  \fi
\fi % End of OFSS
\ifnfssone
  \newmathalphabet{\mathit}
  \addtoversion{normal}{\mathit}{cmr}{m}{it}
  \addtoversion{bold}{\mathit}{cmr}{bx}{it}
  \newmathalphabet{\mathbfit} % math mode version of \textbfit{..}
  \addtoversion{normal}{\mathbfit}{cmr}{bx}{it}
  \addtoversion{bold}{\mathbfit}{cmr}{bx}{it}
  \newmathalphabet{\mathbfss} % math mode version of \textbfss{..}
  \addtoversion{normal}{\mathbfss}{cmss}{bx}{n}
  \addtoversion{bold}{\mathbfss}{cmss}{bx}{n}
  \ifAMStwofonts
    \ifCUPmtlplainloaded \else
      \UseAMStwoboldmath
      \makeatletter
      \new@mathgroup\upmath@group
      \define@mathgroup\mv@normal\upmath@group{eur}{m}{n}
      \define@mathgroup\mv@bold\upmath@group{eur}{b}{n}
      \edef\UPM{\hexnumber\upmath@group}
      \new@mathgroup\amsa@group
      \define@mathgroup\mv@normal\amsa@group{msa}{m}{n}
      \define@mathgroup\mv@bold\amsa@group{msa}{m}{n}
      \edef\AMSa{\hexnumber\amsa@group}
      \makeatother
      \mathchardef\upi="0\UPM19
      \mathchardef\umu="0\UPM16
      \mathchardef\upartial="0\UPM40
      \mathchardef\leqslant="3\AMSa36
      \mathchardef\geqslant="3\AMSa3E
    \fi
  \fi
\fi % End of NFSS release 1

\ifnfsstwo
  \DeclareMathAlphabet{\mathbfit}{OT1}{cmr}{bx}{it}
  \SetMathAlphabet\mathbfit{bold}{OT1}{cmr}{bx}{it}
  \DeclareMathAlphabet{\mathbfss}{OT1}{cmss}{bx}{n}
  \SetMathAlphabet\mathbfss{bold}{OT1}{cmss}{bx}{n}
  \ifAMStwofonts
    \ifCUPmtlplainloaded \else
      \DeclareSymbolFont{UPM}{U}{eur}{m}{n}
      \SetSymbolFont{UPM}{bold}{U}{eur}{b}{n}
      \DeclareSymbolFont{AMSa}{U}{msa}{m}{n}
      \DeclareMathSymbol{\upi}{0}{UPM}{"19}
      \DeclareMathSymbol{\umu}{0}{UPM}{"16}
      \DeclareMathSymbol{\upartial}{0}{UPM}{"40}
      \DeclareMathSymbol{\leqslant}{3}{AMSa}{"36}
      \DeclareMathSymbol{\geqslant}{3}{AMSa}{"3E}
    \fi
  \fi
\fi % End of NFSS release 2

\ifCUPmtlplainloaded \else
  \ifAMStwofonts \else % If no AMS fonts
    \def\upi{\pi}
    \def\umu{\mu}
    \def\upartial{\partial}
  \fi
\fi

\def\lsim{\lower.5ex\hbox{$\; \buildrel < \over \sim \;$}}
\def\gsim{\lower.5ex\hbox{$\; \buildrel > \over \sim \;$}}
\title{The Impact of Net Culture on Mainstream Societies : a Global Analysis}
\author[Tapas Kumar Das]
{Tapas Kumar Das\\
Research Fellow\\
Theoretical Astrophysics Group \\
 S. N. Bose National Centre For Basic Sciences \\
 Block JD Sector III Salt Lake Calcutta 700 091 India\\ 
 E-mail tdas@boson.bose.res.in\\
 Web-Page : http://boson.bose.res.in/$\sim$tdas/home.html}
\begin{document}
\twocolumn
\maketitle
\begin{abstract}

In this work the impact of the Internet culture on standard mainstream
societies has been analyzed. After analytically establishing the fact that the Net can be viewed as a  pan-societal
superstructure which supports its own distinct culture, an ethnographic analysis
is proveded
to find out the key aspects of this culture. The elements of this culture 
which have
an empowering impacts on the standard mainstream societies, as well as the elements 
in it which can cause discouraging social effects 
are then discussed by a global investigation of the present status 
of various fundamental aspects (e,g, education, economics, politics, entertainment etc) of the 
mainstream societies as well as their links with the Net culture. Though immensely potential for
providing various prominent positive impacts, the key findings of this work
indicate that misuse of Internet
can create tremendous harm to the members of the mainstream societies by generating a set of morally
crippled people as well as a future generation completely void of principles and ethics. This 
structured diagnostic approach to the social problems caused by the manhandling of Internet leads
to a concrete effort of providing  the measures that can be taken to enhance or to overcome the 
supporting and limiting effects of the Net culture respectively with the intent to benefit
our society and to protect the teratoidation of certain ethical values.
\end{abstract}
\noindent
% if the twocol comand is here then two col after abstract, i,e, from 
%2nd page
%\twocolumn
\section {Introduction}\footnote{The  {\it superscripts} on some words/set of 
words here give the reference
number in the the {\bf glossary
of selected technical terms} provided at the end of this article (section 10).}
 
Evolving out of the United States ${ARPANET^{1}}$,
first demonstrated in 1972, the
Internet did not begin to approach it's current global size until the mid
1980s. With an estimated user-base of nearly
30
million people in countries as diverse as Australia, Canada, France, Germany,
Italy, India, Japan, Mexico and many other countries, these networks are not
solely a reflection of any one culture, neither do they represent any isolated and
singular society in ethnographic form. Rather the users of these networks 
represent an assortment of diverse cultures as well as a nonstandard homogeneous
mixture of very many societies (societies which explicitly exhibit random
heterogeneity in many aspects) from across the globe. The principal objective of
this project is to determine if a new culture has emerged on these networks, 
distinct from that of the networks' constituent countries. 
If such a culture has emerged, the objective is to 
\begin{itemize}
\item {\bf Provide an ethnographic analysis of this culture}
\end{itemize}
\noindent
and 
\begin{itemize}
\item {\bf Determine what effects (either supporting or limiting) this culture
has on the societies from which ${{newbies}^{9}}$ are being lured into the hypnotic black
hole of the Internet.}
\end{itemize}
\noindent
 Though primarily intended to focus on mainstream Indian society, the 
overall analysis of this work is being pursued in a global context keeping in 
mind the technological ease of communication via Internet between two randomly
chosen points on Cyberatlas, i,e, the concept of formation of an egalitarian
global village is being streaked.
\noindent 
Basically this work will try to provide the answers of the following questions
in a self integrated manner

\begin{itemize}
\item {\bf Do the users of the Net form a society with its own distinct culture?}
\end{itemize}
If so ......,
\begin{itemize}
\item {\bf what are the key aspects of this culture?}
\item {\bf what elements of this culture (if any) have an empowering and what 
elements of this culture  (if any) have a discouraging effect on the standard 
mainstream society carried via the ${{Internauts}^{7}}$?}
\item {\bf what measures can be taken to enhance or overcome the empowering and
discouraging elements of the culture with the intent of easing the process of
enculturation for Internauts to benefit and to prevent the declination of the
ethical values of a mainstream society?}
\end{itemize}
\noindent
 The process of addressing these questions as well as finding their answers 
will be carried out through the proper analysis of present status of various
fundamental aspects of mainstream society (i,e, education, business, 
entertainment, politics etc.) and their link with the Net.
\noindent
\section{Ethnography of the Net culture}
\noindent
 To examine whether the Net forms a distinct society supporting its own
culture, it is necessary to define the terms {\bf Society }and {\bf Culture} in this 
context. A review of the Anthropological literature provides the following most suitable
definitions
\begin{itemize}
\item {{\bf Society:}  {\it  An abstraction of the ways in which interaction among 
humans is patterned}}.({\it Howard(1989)})
\end{itemize}
In layman's language the definition could be...
\begin{itemize}
\item {\it  A more or less organized group of people of both sexes that share a 
common culture}.({\it Barnouw(1987)})
\item {\bf Culture:}  {\it A culture is the way of life of a group of people, 
the complex of shared concepts and patterns of learned behavior that are 
handed down from one generation to the next through the means of language and
intuition}.( {\it Bernard(1988)} )
\end{itemize}
\noindent
There are other different ways to 
narrate what society and culture are. Without giving details about other 
definitions, it can be said that, though not all in the same fashion, all
definitions by and large make a clear distinction between the concepts of 
society and culture. What then are the dependencies between society and culture?
Can a society have more than one culture or can the same culture be shared by 
It seems quite 
reasonable to assume that all the members ({\it having equal financial background})
of a particular society 
by and large share almost the same culture. It is 
also implied in the definitions of culture presented by {\it Ember
and Ember(1990)}
and 
{\it Nanda(1991)} that two different societies cannot posses exactly same culture.
One might concede that there is a small possibility that two unrelated societies
may, perhaps as a result of similar environment, independently evolve identical
culture, but in practice it can be argued that usually this does not happen.\\[0.25cm]
\noindent
The above discussion is significant in this context because together 
they imply that if the users of the Net are to be described as having a 
distinct culture then they need to form one particular separate society. It would
appear from the global span of the Net that this is not the case. This
dilemma can be resolved by viewing the Net as a {\it pan-societal
superstructure} ({\it North(1994)})  which is freed of the responsibilities of providing a number of
properties that can reasonably be expected from any mainstream society (e,g, 
reproduction, food and shelter) by virtue of the fact that its members are also
the members of traditional mainstream societies that {\it do supply} all these
features. Rather it is a melting pot of different components (such as 
socialization, economics, politics, entertainment) of many of the standard 
mainstream societies. Thus the Net {\it should be viewed not as an independent
society} (for it does not provide all of the features of an independent society)
, but as a superstructural society which, in a statistical view, can be
described as the {\it superset of intersection where the overlapping subsets are
the distinct mainstream societies}.\\[0.25cm] 
\noindent
Regarding the Net's culture we may say this;
 though it does share many elements with the cultures of the mainstream
societies 
that it spans, much of it is taken from 
these mainstream cultures, there is also much in it that represents culture
adaptization to a new environment. Now if the question arises whether the Net
might be considered to be a subculture of some other culture of any 
other particular mainstream society, it is reasonable to
ask {\it What could the Net be a subculture of?} Given its global 
spanning size, it is obvious that
there is no {\it one} society or culture that can subsume the Net.
It therefore seems ineffectual to think of it in subcultural terms. So while
inheriting many cultural elements from the societies that it spans, {\it this 
superstructural Net society nonetheless  supports its own distinct culture}.
\noindent
\section{Analysis of the Net culture}
\noindent
A survey of different literature and a thorough netsurfing gives the following
key aspects of the Net culture.....
\begin{itemize}
\item {Written adaptations to the text-only medium 
for user to user communication include the use of 
${{emoticons}^{5}}$
 to denote emotional intent and the use of characters such as 
``*" and ``-" to denote emphasis.}
\end{itemize}
 {\bf Conclusion: Language of communication gets a new shape?}
\begin{itemize}
\item {In a mainstream society many factors determine how we judge and are 
judged by other people. Such factors include appearance, gender, race, wealth,
 dress and occupation. On the mainly text-only channels of intracommunications
 via Net, these factors are difficult to determine and one is left with far 
fewer criteria on which Nauts are being judged. Prestige is acquired within 
the Net culture primaryly through what one writes or through philanthropic
actions such as maintaining a mailing list or writing freely distributed
softwares etc.}
\end{itemize}
{\bf Conclusion: A new concept of acquiring social prestige, apparently
more healthy than
conventional one, is launched.}
\begin{itemize}
\item On the contrary to the previous feature, for those individuals who are
relying on being judged or to judge by such factors as their appearance, wealth,
sex appeal etc, the mainly text only communication base
of the Net may be an uncomfortable
experience.
\end{itemize}
\noindent
{\bf Conclusion: Counter-culturists on Net?}
\noindent
\footnote
{Though the introduction of various highly powerful and user-friendly graphics
packages changes the text only nature of communication day by day, still there are scopes
for them who don't want to get exposed about their features. Different Net-friendship, Net-dating
sites are the proper examples for this purpose.}
\begin{itemize}
\item {The increasing size of the Net leads to an increased opportunity for the 
exchange of various resources. Resources such as advice, graphics and softwares are
most often handed on the Net through a reciprocal exchange system. With the 
increasing size of the Net yielding a higher population of users available to 
supply resources, it follows that there is a greater pool of resources from which 
to draw information. This is a curiously inverted position to that of a 
mainstream society. In such a society, when resources is shared amongst a number 
of people, the greater the population the more thinly the resource is distributed.
On the Net, however, if a resource is made available for ${{FTP}^{6}}$ then it can be copied
by any number of people without lessening the share of the resource available to 
each.}
\end{itemize}
\noindent
{\bf Conclusion: Reciprocity is an important feature of the Net's economy. 
Signature of symbiotic culturization?}
\begin{itemize}
\item {Social stratification is present within the Net society; for example, 
system administrators and news group moderators have powers that most users do not.
The most vital point in this regard (which will be discussed in detail later on) 
is, due to financial, educational and technical constraints, not all users
 have equal access to the resources.}
\end{itemize}
\noindent
{\bf Conclusion: Net capitalism?}
\begin{itemize}
\item {The Net is viewed by some of its misusers as an anarchy which can cause 
(and in reality, {\it it  does}
 cause) severe moral degradation of the members (mainly, of the teen-agers) of the 
mainstream societies in a direct or indirect way. (Detail discussion will be 
presented in the next section)} 
\end{itemize}
\noindent
{\bf Conclusion: Degredation of ethical values through Net. }
\begin{itemize}
\item {However, the Net culture has a complex set of conventions and lore called
{\it Netiquette} to which users are expected to conform. Transgressions of which
are supposed to be dealt by other users in various ways ranging from written
chastisement to the invocation of police authorities of mainstream societies.}
\end{itemize}
\noindent
{\bf Conclusion: Protection of teratoidation of Net societies?}\\[0.25cm]
\noindent
After these general discussions about the nature of the Net society and its culture,
it is now quite reasonable to fragmentize the analysis into diverse social 
issues (like education, business, politics etc) case by case and to study the 
correlation between the mainstream societies and the Net in these contexts.
\noindent
\section{Correlation between the mainstream societies and the Net}
\noindent
\subsection{Education}
\noindent 
As already been discussed, Nauts have a facility to hide their racial, financial
and even generic identity if they want. This makes the Net as a
medium of distant education and learning. Net has immense power to change
the whole infrastructure of educational system in a highly positive way. According
to Neil Rudenstein, a famous academician from Harvard University, USA, education
is basically a dynamic and continuous process which can give it's best 
output if the
student-teacher interaction can be maximized. The chemistry of this interaction
is the backbone of a proper education system. In our mainstream societies,
orthodox educational infrastructure is insufficient to provide the maximization
of this interaction. All students do not get equal time to interact with the 
teacher nor they are being paid equal attention. There also 
remain some psychological
barriers which prevent the mobilization of the dynamism of the teacher 
student interaction process, for example, the shyness of a student to ask questions 
in the class or the inferiority complex generated by social or financial backwardness.
These problems can be overcome if the Net can properly used as an educational 
media because academic interactions via the Net will make the teacher-student 
interaction more effective and vibrant. According to Nancy Singer, a Prof. 
of Colorado University, Department of education, application of the Net as an 
educational media helps students to overcome shyness and to enhance their power of
communication one example of which is the publication of the world's
first {\it fully }net-operated newspaper maintained by the teen-ager 
students of the Centeniel school.
Moreover, the Net's voluminous information
bank, along with the application of different graphics packages and audio clips makes the 
education more interesting and effective.\\[0.25cm]
\noindent
Here one crucial point comes. What does education means really?
When should a person
be called  educated? Does the process of education mean only accumulation
of information and data and to a make career? The answer 
is  a strong  NO!
Real education helps to grow certain sense of principle
and ethics. Without them education is incomplete. It can provide a person 
a very good career and wealth but can never make him/her a proper human being.
And as all these value-related issues of the education have concrete dependence 
on the ethical views of the teacher concerned, Net can {\it never} give 
complete 
education independently. Regarding communication and information gathering, Net
has absolutely no parallel, but its control {\it has to be} 
taken over by human being when question of generation of values arises.
\noindent
\subsection{Spreading of harmful materials}
\noindent
As already mentioned, some of the misusers of the Net is viewing it as an
anarchy. The entry of any arbitrary user in Net world has created many crucial 
social problems all of which are carried to all the mainstream societies via 
the Net 
superstructure. The most fatal problem is the {\it cyberporno} - the open 
business of pornography in the whole realm of the Net. As there 
is no bar in entry into the Net, companies like Brandy, Web and Penthouse have 
launched their nasty business of pornography throughout the Net. A recent report
from a research group of Carnegi Mealon University, USA, shows that 83.5 percent of 
the stored pictures in Usenet are pornographic in nature. Three years 
back, in 1995, US Government launched the {\it Communication Decency Act}
({\it CDA})
 but it is a pity that nobody could be arrested under this act because almost 
all the companies of USA made appeal in the Supreme Court against this act 
claiming that it violates their democratic freedom of speech. In 1996, the
German 
Government has declared some of the sites as illegal and entry into those sites
are kept prohibited thereof. Recently the Singapore Government also had to 
take the same 
step. Smith System Engineering, a UK based consultancy, has been awarded a contract from
the European parliament to investigate the fearability of jamming pornography
and racism (another fatal problem created by the racial group of people such
as Neo-Nazis) on the Net. The project, expected to to run for six months, will
examine methods where by offensive materials are distributed and study the
technical method of blocking their flow.
Unfortunately all these processes could not make the situation that much 
better. Millions of the {\it Net-porno} web sites are there entry into which
requires only an {\it Adult check ID password} which can easily be accessed 
by paying only 15 - 20 US ${\$}$ and as payment is done by credit cards 
via Net, 
there is absolutely no way to check whether the applicants for this card are 
really over 18 by age or not! So millions of teen agers are accessing these 
dirty sites regularly which can result (and practically, which {\it does} result)
 a severe degradation of ethical and moral values worldwide. In western capitalist 
countries, such as in the USA and some countries from central and western Europe, 
numerous teen-age crimes have been reported to police which, according to the 
modern sociologists, are believed to be committed from odd sexual obsessions
generated by different pornographic sites of the Net. Not only in western
countries, Asia's rapid growing economies are using the Net to 
integrate themselves with the world economy but the governments of some of the 
regions are not entirely comfortable with the phenomenon. The rise in the 
popularity of the global networks has alarmed some governments which fear 
unbridled access could lead to distribution of pornography, spread religious
unorthodoxy or encourage political dissidence. None of them has come up with 
any foolproof plan to screen undesirable materials on Net, and experts say 
such control is technically very difficult.\\[0.25cm]
\noindent
So it comes out that ultimately not the technology, but our ethical
sense 
matters to make the society free from evils of misuse of Net. The combined
education, taste, culture, principle and ethics of the Nauts  are the recipes
which can make the Net culture clean, and {\it here humanity takes over 
technology}.
\noindent
\subsection{Business and Industry}
\noindent
Regarding business, the Net has brought massive changes in the way of business in 
financially forward countries. Small companies have got a golden 
opportunity to spread their business in international market through the 
commercial home pages and web sites. Introduction of Net helped the Software
companies to make maximum profit. Netscape Communication, an USA based software
making company first launched the Netscape Navigation Browser which made 
Netsurfing very easy. After that, Microsoft and Oracle made different softwares
which nowadays have immense application in Netsurfing. Besides this, Digital 
Equipment and Sysco Systems are involved in data transfer technology via Net.
The most important contribution came from SUN Microsystems which introduced 
JAVA, a very powerful language for different kind of Computer applications in 
the Net. On the other hand, the wide spanning of the Networks has created 
problems for some computer companies also. As for the example, the future market
of the softwares {\it Workgroup} and {\it Networking} developed by Novel and Lotus
(two branches of world famous IBM group) became uncertain.
Another important contribution of Net is the introduction of a couple of very
powerful {\it Online Information Services} e,g, NETCOM, UUS, PIPEX, DAEMON etc.
 Their wide application has lessened the business of electronic world (e-world)
package of some pre Net online services, such as of Microsoft and Apple.\\[0.25cm]
\noindent
Probably the most stirring social effect of the introduction of these online
information services awaiting in the near future
is the sharp blow on paper and printing industries. Heavily
graphics and audio clip supported www is fostering the commercial use of the Net
by making it easy and fun for Nauts to check out informations and products
whenever they wish with as much depth of multimedia detail as they want. News-
papers, books, magazines and journals can directly be accessed from the Net more
easily and quickly than in printed form. 
This may cause a decreasing want of printed 
materials in near future resulting the deterioration of the condition of the paper and printing
industries. Who will provide a parallel job for the workers in these industries
has become a major question.
Not only in paper and printing industries, the online shopping system also 
may create such kind of problems. Worlds largest retailer Wall Mart, UK's famous
Berkley Group and other different companies have launched their shopping malls
in Net which results in the unemployment of regular workers in various shopping
complexes.
\noindent
\section{Indian perspective}
\noindent
Focusing the analysis on the Indian mainstream society, it is quite reasonable to
say that situations differ in India compared to the western countries, 
the reason for this will be clear from the discussion hereafter.\\[0.25cm]
\noindent
The rate of literacy in India is around 52 $\%$, among them, the number of 
persons with a strong background in English is countably small. So a very small fraction of
950 million Indians have academic qualification to deal with the
Net. Besides this, one needs around 50 Thousand Rs/$\_$ for bying a computer along with the 
standard VSNL charge for getting an Internet connection. Only a small part of 
the literate Indian population can afford this amount. So the percentage of 
the total Indian
as private Net users is very small. In our country, Net facilities are accessed
basically by people involved in academic world, e,g, in Universities and
research institutions. But the rate at which the market price of computer
and electronic components are decreasing may sharply change the situation in near
future. 
However, the Net consciousness is continuously increasing. 
Here the view differs among the common people. Here, though a part of the Net
users are thinking to use the Net to enhance the academic standard in our
country, the major part of the Net using community are welcoming Net 
looking for their own business interests. The {\it Internet Service 
Providers} ({\it ISP}) are seeing the Indian market very much promising as well as 
virgin after the {\it Open Market Policy} launched by Indian Government. 
Here one crucial point must be emphasized, in a poor but promising 
country like India, all the policies regarding the use and spread of
Internet should be made very carefully because  dealing with the 
interest of common people is an extremely sensitive and delicate issue in our
country. India has its own completely different form of societal
infrastructure compared to the western countries and even than the other 
developed Asian countries also. The form of the social, political and 
economic problems of the Indian mainstream society is an absolutely disjoint
set. Here in our country the fundamental accesses and facilities to the 
essential need are very thin. Under these circumstances, determining the policy 
which will fix up the usage of the Net needs  very special attention as well
as  deep thought because the prime aim of spreading the Net in a third world
country should necessarily be focussed on the issue of the development 
and prosperity of common mass;
like all other facilities, it also should not
be unevenly heaped onto a small class of capitalist elements. The policy-makers
should not be biased to lump all the facilities of the Net only among the upper class
 keeping in mind the fact that mass communication is a vital factor for 
the overall development of an underdeveloped third world country and the Net has
immense potential for this purpose. The spreading of the application of Net
amongst the have nots of Indian society should definitely bring a prosperous
future if the supreme control is in proper hands anyway.\\[0.25cm]
\noindent 
A spark of effort in this regard, though very tiny, (still which shows a very
bright prospect) is being observed recently in Andhrapradesh. Ranadeep Sudan, an additional
secretary of Chandrababu ministry has made a rough sketch about how to use the
Net as the mass communicator between the various public sectors and the 
common people. This is being considered by the experts as a very important
step on the Government policy of Information technology. In this  concept 
paper of using the Net properly, it has been explained thoroughly how, by 
2001 AD, electronic network will start operating fully to communicate 
between the administration and the common mass. All sorts of official informations,
application forms for jobs, electricity, water, and even for the individual
ration cards will be available online and instantly for 24 hours via Net. 
Parallel to the public sector administrations, hospitals, nursing homes, 
libraries and all other voluntary services and organizations will be connected
through the Net. For the convenience of the common people, intermediate Internet 
booths (like the local telephone exchanges) will be operated which will 
communicate the vital nodes throughout the state resulting smooth and fast
operation of the Net. Clearly the vision behind this effort is absolutely 
magnificent. This will lessen the hazards encountered in various
public sectors while dealing with the problems of common people and to 
lubricate the dynamism of activity in the administration and secretariat.\\[0.25cm]
\noindent
Parallel to these kind of individual philanthropic activities, the open market
policy in our country puts the role of VSNL (Videsh Sanchar Nigam Limited) 
as a questionable issue. According to the present status of the Central 
Government policy, only VSNL will moderate all sort of Internet connections
 (institutional as well as private), upto 2004 AD. This tenure may be 
stretched indefinitely in future. The monopoly of VSNL as the gateway access
has become a contentious issue among the people who are inclined towards the 
privatization of the Net administration in India.
\noindent
\section{The Information Fatigue Syndrome: a new socio-psychological problem}
\noindent
One striking social problem worldwide 
generated by the Internet use is {\it Internet 
Addiction.} Which, though didn't directly affect Indian mainstream society
that much till now, has been widely spread among the Net users in western 
countries. Internet addicts are said to spend increased amount of time
online and feel agitated or irritable when off-line. Even without a 
key-board at hand, a few make voluntary or involuntary typing movements with
their fingers. Many of them have practically withdrawn themselves from mainstream 
social activities. Some find themselves with around 500 US ${\$}$  
monthly
online bills which they can't afford. Some of the modern counsellars
(psychoanalyst) found by extensive research on a broad group of people that
excessive use of Internet use can have serious consequences, from mental 
depression to dropping out from school or college, {\it even divorce}.
Nowadays, in most of the western countries, the Internet addiction has been
compared to the drug addiction. Being a ${behavioral}$ ${{addiction}_{2}}$,
the Net addiction has been closely compared to the {\it marijuana}
because the marijuana plant Cannabis Sativa contains a group of chemicals
called {\it tetrahydrocannabinols} ({\it thcb}) which activates the stimuli
that is activated almost in a same way by heavy Internet use. 
However, the {\it thcb} group can have 
more direct physiological effects than the Internet, like major 
physiological effects on the cardiovascular and central nervous system,
(note that the Net can also have effects on the central nervous system,
though smaller in extent compared to the {\it thcb} group) short term loss
of memory, loss of balance and difficulty in completing of thought
processes.  According 
to one eminent psychiatrist, {\it Internet is a strong psychostimulant}.
Internet addiction, like any other addiction, has sign and symptoms which can
be recognized by the addict and by those close to the addicts. This addiction
is technically termed as {\it Information Fatigue Syndrome}. 
Some of the possible physiological correlates of heavy
Internet usage may be listed in following manner...
\begin{itemize}
\item {\bf A conditional response( increased pulse, blood pressure) to the 
modem connecting.}
\item {\bf An {\large\it altered state of consciousness} during long period of 
${{dyads}^{4}}$/
 small group interaction (total focus and concentration on screen, similar
to a meditation/trance state)}.
\item {\bf Dreams that appeared in scrolling texts and pictures.}
\item {\bf Extreme irritability when interrupted by people/things in 
real life while immersed in cyber-space.}
\end{itemize}
\noindent
Like all other addictions, whether a chemical addiction or another type, Net
addiction also has some strong side effects on mainstream societies. The
Net addiction makes users to ignore their surroundings as well as to 
withdraw themselves from the activities of the mainstream societies, they even
intend to commit crimes when their online activities are interrupted by 
somebody. ${Chat}$ ${{rooms}^{3}}$ 
are said to be the most addictive aspects of the 
Internet. According to Dr. Howard Shaffer, the associate director of the 
division of addiction at Harvard University and Dr.Ivan Goldberg, MD 
Psychiatry of the 
{\it Internet addiction support Group}, the most fatal and widespread effects 
of the Internet addiction to the Net users is abruptly 
reduced (or even completely given up
in many cases) important occupational, interactive and
recreational activities  because of excessive
use of the Net which, in turn, affects the mainstream society resulting the 
complete demolition of well-organized intrasocietal interaction pattern.\\[0.25cm]
\noindent
Internet addiction is considered to be a serious social problem where 
treatments and remedies need to be sincerely found out. The most prominent 
effort in this aspect is the {\it Centre For On-line Addiction} ({\it COLA}) 
 which is
a {\it cyberclinic} devoted to study the cyberspace addiction headed by Dr. Kimberly S. Young.
Some other voluntary online cyberclinics are also established to fight against
the Net addiction among which the most promising one is the {\it Internet
Addiction disorder support Group} by Dr. Ivan Goldberg.
\noindent
\section{Future perspective (global)}
\noindent
30 years back, Telecommunication maestro Marshall Macluhan launched the dream
of making {\it global village} ; which is not far from reality today with 
the gigantically powerful cybercommunication system in hand - the
Internet. It should not sound over exaggerated if the Internet is called the
{\it most powerful and the largest machine ever constructed by human race}. 
With around 50 thousand networks, 6.6 million computers and 30 million daily 
users, this {\it information super-highway} has linked 160 countries of the
world. In 1993, The Clinton administration of USA 
made a very important administrative 
planning - {\it National Information Infrastructure} ({\it NII})  the 
aim of which was to 
create a {\it network of computer networks} which will allow the integration
of computer hardware, software, telecommunications and shells that will make 
it easy to connect people with one another via computers and the access 
towards a vast array of services and information will be widely open by this
effect. A proper application of this will lead us to the concept of 
{\it Global Information Infrastructure} ({\it GII})  under the operation of which the
the remotest part of the globe can be connected to any other place
of any country within a fraction of a second. In a recent interview given
to The Newsweek magazine, 
Prof. Russell Newmann (head, department of 
International communication, Kennedy School of Government, Harvard) said that 
the future global village, independent of the geographical and political
barrier, would be built on the basis of mentality of the members 
of different 
mainstream societies.
 People having same or almost same mentality and way of
thinking will be grouped together to make a united, clean and egalitarian 
new society with its own culture adopted from, and given back to the Net culture.
 But will this dream of equality among the people be fulfilled? Will a person
from a poverty stricken third world country get equal opportunity and facilities 
in comparison to a Net user from Western capitalist countries having a strong
financial background (which usually makes all the differences in mainstream 
societies) in that global village? Or, more precisely, do the people coming from
different financial and social background {\large\it really}  acquire equal
prestige in the Net culture? Does the Net culture support that much democracy?
If so, who will determine the ultimate fate of the human race, is it the class
of {\it couch potatoes}, sitting lazily and having the computer mouse in hand
, who will dictate the future of mankind? The spreading of Net across the globe
at exploding rate makes the whole human race to face all these very very crucial questions. Effort will be made to provide the answers of these questions
in next sections.
\noindent
\section{Quo Vadis ?}
\noindent
The whole analysis of this work reflects the idea that the structure of the
mainstream societies as well as the activities of its members {\it do get 
influenced very much by the pan-societal superstructure of the Internet and by
the culture evolved in it}. This invites the following most fundamental
questions....
\begin{itemize}
\item {\bf Does the Net society and its culture support that much 
democracy which can make us dream to have a future society where
everybody will get equal share of all fundamental rights in a 
socialistic infrastructure?}
\end{itemize} 
\noindent
if not, then......
\begin{itemize}
\item {\bf Will the Net culture, along with its social stratification, lead us 
to a future world where various essential resources will be accumulated only
among the people having strong financial background, e,g, will the global village  
become a capitalism dominated society?}
\end{itemize}
\noindent
and in the worst case.....
\begin{itemize}
\item {\bf Does the Net culture contain that many elements of anarchism which
can demolish all the ethical values in future mainstream societies
as well as in the global village by creating an ethically crippled future 
generation void of principles and values?}
\end{itemize}
\noindent
To draw the final conclusion by searching the answers of the questions addressed
above, it is necessary to have a critical eye on the classification of the 
mainstream societies in {\it anthropological, economic and political viewpoint}
and to the structured correlation of the Net culture with it. \\[0.25cm]
\noindent
One important way in which anthropologists classify different societies is {\it
according to the degree to which different groups within a society have unequal
access to the advantages such as resources, prestige and power} 
({\it Ember and Ember(1990)
, Murphy(1989), Nanda(1991), Howard(1989)}). 
The stratification of different groups within a 
society gives rise to three different type of societies being generally recognized......
\begin{enumerate}
\item {\bf Egalitarian Society: } {\it Societies with the least stratification}. 
\item {\bf Ranked Society: } {\it People are divided into hierarchically 
ordered groups that differ in terms of social prestige, but not significantly
in terms of accesses to resources of power.}
\item {\bf Class-based Society:} {\it People are divided into hierarchically
ordered groups that differ in terms of access not only to prestige, but also
{\it and mainly to} resources and power.}
\end{enumerate}
\noindent
However, though the reciprocity of exchanging various resources and the new
concept of acquiring prestige (already discussed at the beginning) apparently
represents the {\it semi-egalitarian} features of the pan-societal Net 
superstructure, the individual wealth necessary to support the cost of the 
Net access via a commercial carrier limits the number of people from mainstream
societies to access the facilities of the Net which clearly indicates the 
presence of strong capitalism on Net culture. It is quite obvious from the 
argument given above that {until and unless Government from all the 
mainstream societies can provide equal financial facility to all the members 
of the society for accessing the Net, which is far from reality in approximately
all the class-based societies, instead of being egalitarian one, the future
global village will obviously become a strongly capitalism dominated 
society.} So it seems that the {\it only way } to make the global village strongly
egalitarian is  the establishment of socialism in all the mainstream societies.
Regarding the Net culture, because supporting advanced computer technology
is strongly involved with the question of national wealth, it is quite
obvious that for all the mainstream societies, percentage right as well as
the scope to contribute their individual cultures to the melting pot of the
grand Net culture will depend on their society wise financial strength inequality 
of which clearly indicates that the Net culture also will be dominated by the
members of the financially onward societies only. In such a culture, though 
apparently the process of acquiring prestige is newly defined, ultimately individual,
as well as the national wealth will be the only dominant factor.\\[0.25cm]
\noindent
Along with the above mentioned complexities, there is a strong possibility 
that  Net culture may deteriorate the ethical values in the future 
generation of the mainstream societies worldwide by some of  
the anarchical activities of its misusers. Though the absence of a centralized decision-making body
and its rule-of-the-people approach of the Net seems, at first, to accord 
classical democracy a good likeness to the political structure of the Net, 
however, the approach of the classical orthodox democracy is {\it not} a 
good metaphor for the political structure of the Net because in practice, 
democracy is best seen as a principle involving political consent and control
on the part the governed that may find expressions in various political 
practices and forms of government, which, in practice, is absent in the Net
culture. Rather the Net is viewed as an anarchy by many of its members,
though this anarchism is not as formal as the nineteenth century definitions
of anarchism described by {\it Michael Bakunin} and {\it Peter Kropotkin} essence of 
which are  ({\it Hagopian(1985)}) .....
\begin{itemize}
\item {\bf The negation of state.}
\item {\bf The abolition of private properties.}
\item {\bf Revolution.}
\item {\bf The rejection of religion.}
\item {\bf The foundation of a new co-operative order.}
\end{itemize}
\noindent
Judging from the revolutionary tone of the above list, apparently it seems quite incompatible
to describe the Net culture as an anarchy. As for example, till today no claim
has been made that because the Net society is an anarchy, it should advocate 
the rejection of religion. (Rather there are many websites which configure 
profound persuasion in favour of different religions of vary many diverse
mainstream societies.) So what then does it mean if the Net culture is 
described as an anarchy? The following section provides insight into it.\\
\noindent
The term anarchy comes from {\it anarch}, which means {\it without leader}. Taking
the definition of anarchy as a society with no government that policies itself
by peer review, and with no laws imposed by central authority, the Net 
perfectly fits the bill as the anarchist's philosophy of individual
responsibility is completely fulfilled in this culture. Though here the term
{\it individual} may not always indicate the individuality of a single
user rather it points out mostly to the individual websites. In general, the
system administrator or the webmaster is the absolute ruler of his site which,
in fact makes the sites a {\it dictatorship rather than anarchy} when 
considered individually. (This is not necessarily bad always because many are
benevolent dictatorship.) At the same time there is no possibility where
system administrator (or webmaster) of a particular site can set the rule
for any other site and there is absolutely no concept of any {\it super
administrator} which will govern all the individual sites of the Net. So the
Net is {\it not} centrally governed although the individual sites are.
Thus even though {\it locally} the individual sites may not, the Net as a 
whole can really be described as a {\it co-operative anarchy} which carries the following features of anarchism... 
\begin{itemize}
\item {\bf Absence of government.}
\item {\bf No form of centrally imposed control.}
\item {\bf No mechanism available for correction to take place at a Net-wide
level.}
\end{itemize}
\noindent
Another very uncomfortable feature of the Net culture is the demolition
of identity and ethnic heritage of a classical society. {\it Every 
member, every information, every individual deep thought is nothing but 
just a number, a website, an ${IP}$ 
${{address}_{8}}$ in the Net culture.}
\noindent
\section{Epilouge}
\noindent
So ultimately it seems that though the Net culture has prominent positive 
effects on mainstream societies, {\it central control of it should always be 
in proper hands} because {\it computers can never take over human being 
as long as the question of ethical sense is concerned.} We thus always should 
keep in mind that the Internet, being immensely powerful to change the 
whole structure of the human society, {\it should be handled 
very very carefully with the intent of philanthropic activities.} Internet, 
 the {\it most crowning achievement} from the present technology, can be 
described as a sharp weapon. Whether the use of this weapon should be as a 
surgical knife to operate the evils of the society or as a killing chopper, 
that decision is left to the humanism of its users as well as to the future
generation.
\section{Glossary of selected technical terms}
\begin{enumerate}
\item {\bf ARPANET :} The world's first computer network was a U.S. government-funded wide area network
called the ARPANET ({\it Advanced Research Project Agency Network}). The first ARPANET {\it Information
Message Processor} (IMP) was installed on Sept 1, 1969 which had only 12 KB of memory although they were
considered to be powerful minicomputers of their time. In the early 1980's the ARPANET was split 
into two separate networks, the new counterpart was the MILNET, a non-classified US military network
which is still in service having it's headquarter in Virginia. For an excellent overview of the 
history of the Net, see\\
\noindent
{\bf The History Of The Net\\
\noindent
Master's Thesis by Henry Edward Hardy\\
\noindent
School Of Communications\\
\noindent
Grand Valley State University}
\item {\bf Behavioral Addiction :} A behavioral addiction is one in which an individual is addicted to
an action and {\it not necessarily} a substance. Here people may become addicted to activities even 
when there is no true physiological dependence. 
\item {\bf Chat Room :} Users within the Net community electronically interact with each other
through {\it chat} programs which provides a textual interaction amongst its users. Participants
do not speak to each other, rather they type their {\it dialogue} at their computer key board which 
then appear on the screens of all the other users on-line. Websites devoted to this business is called
{\it chat rooms}.
\item {\bf Dyads :} Online interaction between {\it two} persons via computer.
\item {\bf Emoticons :} The lack of verbal and visual cues in the textual medium of the Net has 
resulted in its users developing and adopting symbols (which {\it can } be typed by key board) 
to clarify their emotional intent. Those symbols are called {\it emoticons}. For detail informations
and a complete list of the emoticons, see\\
\noindent
{\bf The New Hacker's Dictionary\\
\noindent
Raymond, E. ed (1993)\\
\noindent
2nd edn. Cambridge: MIT Press}
\item {\bf FTP :} The Net enables its users to copy files from different other computers on the Net and their own
keeping the original source file intact. This process is called {\it file transfer} and the program that
is used for this purpose is called {\it File Transfer Protocol} ({\it FTP}).
\item {\bf Internauts :} The person who surfs through the Internet.
\item {\bf IP Address :} Any computer connected in the Net has its own and {\it unique} number (like
206.101.167.195) which is used to identify or to refer that computer in the Network System. 
\item {\bf Newbies :} A person who surfs through different websites are named as {\it webies} so new webies
are referred by this term.
\end{enumerate}


\begin{thebibliography}{}
\bibitem[] {} {\bf Barnouw, V. (1987) : } {\it An introduction to anthropology: ethnology.} 5th ed. Chicago: The Dorsey Press. 
\bibitem[] {} {\bf Bernard, H. A. (1988) :} {\it Research methods in cultural anthropology.} Newbury Park, California: saga publications.
American, 265(3): 30 - 41.
\bibitem[] {} {\bf Ember, C. R. and Ember, M. (1990) :} {\it Cultural anthropology. } (6th ed.) New Jersey: Prentice Hall.
\bibitem[] {} {\bf Hagopian, M. N. (1985) :} {\it Ideals and ideologies of modern politics. } New York: Longman.
\bibitem[] {} {\bf Harris, M. (1988) :} {\it Culture, people, nature: An introduction to general
anthropology. } 5th ed. New York: Harper and Row.
\bibitem[] {} {\bf Howard, M. C. (1989) :} {\it Contemporary cultural anthropology. } 3rd ed. Glenview, Illinois: Scott, Foresman and co.
\bibitem[] {} {\bf Murphy, R. F. (1989) : } {\it Cultural and social anthropology. } 3rd ed. Englewood-Cliffs, New Jersey: Prentice Hall.
\bibitem[] {} {\bf Nanda, S. (1991) :} {\it Cultural anthropology. } 4th ed. Belmont, 
California: Wadsworth. 
\bibitem[] {} {\bf North, T. (1994) :} {\it The Internet and Usenet Global Computer
Networks.} A masters thesis.
\end{thebibliography}
\end{document}